\documentclass[10pt]{IEEEtran}

\usepackage[letterpaper,hmargin=1.0in,vmargin=1.0in]{geometry}
\usepackage{graphicx}
\usepackage{subfigure}
\usepackage{listings}

\newcommand{\flaim}{FLAIM }
\begin{document}
\centerfigcaptionstrue
\date{}
\title{\Large{\bf FLAIM: A Multi-level Anonymization Framework for
    Computer and Network Logs}}
\author{Adam Slagell \ \ \ \ \ \ \ \ Kiran Lakkaraju \ \ \ \ \ \ \ \ Katherine Luo\\ \\National Center for Supercomputing Applications (NCSA) \\ University of Illinois at Urbana-Champaign\\ \{{\it slagell,kiran,xluo1}\}{\it @ncsa.uiuc.edu}}
\maketitle

\begin{abstract}
\flaim (Framework for Log Anonymization and Information Management)
addresses two important needs not well addressed by current log
anonymizers.  First, it is extremely modular and not tied to the
specific log being anonymized.  Second, it supports multi-level
anonymization, allowing system administrators to make fine-grained
trade-offs between information loss and privacy/security concerns. In
this paper, we examine anonymization solutions to date and note the
above limitations in each. We further describe how \flaim addresses
these problems, and we describe \flaim's architecture and features in
detail.
\end{abstract}
\section{Introduction}\label{intro}
As computer systems have become more interconnected and attacks have
grown broader in scope, forensic investigations of computer security
incidents have crossed more and more organizational boundaries
\cite{Markoff05}. This poses a difficulty for the computer security
engineer since it becomes more difficult to understand attacks from
the narrowing perspective they have from the vantage point of just
their own logs.  Consequently, there is an increased desire to share
logs within the security operations community.

This increased demand is clearly seen within the community of security
operations, but developers, researchers and educators also depend upon
log sharing. Developers of forensic and log analysis tools need
records from real incidents to test the effectiveness of their new
tools. Networking researchers depend heavily upon large and diverse
data sets of network traces \cite{CAIDA}.  Security and honeynet
researchers also desire large and diverse data sets of logs and
network traces to do their research \cite{Vrable05}. Educators in
traditional academia as well as those that train security engineers
(e.g., the SANS Institute) depend upon real-life examples for students
to analyze and incorporate into assignments. Consequently, in step
with the growth of the computer security field, there has been an
increase in the need for sound methods of sharing log data.

While it is generally recognized that sharing logs is important and
useful \cite{Slagell05}, it is very difficult to accomplish even among
small groups. The difficulty arises because logs are sensitive, and it
is difficult to establish high levels of trust between multiple
organizations---especially in a rapid manner in response to
distributed, yet related attacks. The consequences of logs getting
into the wrong hands can be severe whether they are simply mishandled
by friendly peers or fall directly into the hands of adversaries
through public disclosure.

There are many types of potentially sensitive information in logs, and
as such, logs are like a treasure map for would-be attackers. Access
to them can provide special views of security weaknesses not visible
from outside scans, or in the very least make the attacker's
reconnaissance job much simpler. Logs may reveal bottlenecks as
potential DoS targets, record plaintext login credentials, reveal
security relationships between objects or reveal soft targets (e.g.,
machines already compromised and part of an existing botnet). For
these reasons, security administrators are reluctant to share their
logs with other organizations, even if there is a potential benefit.

While a draconian vetting process for would-be recipients and strong
physical controls over the location of log data could be used to
address some of these concerns about sharing sensitive logs, such 
techniques are not flexible or scalable. Anonymization is an eminently
more flexible solution, and it has been increasingly employed in
recent years. Still, the tools and methods to-date that anonymize logs
are immature, and log sharing has not been achieved at the levels
expected or with the ease desired.  It is true that there are log
repositories out there, but they all seem to share at least one of the
following problems: (1) They do not have a wide view of the Internet,
but they are quite localized, (2) the repositories are very specific,
addressing only one or a few types of logs, (3) anonymization
techniques are weak---often nonexistent---and usually inconsistent
between submissions, or (4) they collect many logs but do not share
them with the research community \cite{Slagell05}. Furthermore, even
if someone is willing to take a risk and submit a large number of logs
for public consumption, security engineers are still faced with the
difficulties of gathering specific logs from other organizations
during an on-going investigation.

We contend that the problem stems from the fact that tools for log
anonymization are immature and not able to meet the many, varied needs
of potential users. Specifically, they have been one-size-fits-all
tools, addressing only one type of log, often anonymizing only one
field in one way. We desperately need more flexible tools that are (1)
highly extensible; (2) multi-level, supporting many options for each
field, allowing one to customize the level of anonymization for their
needs; (3) multi-log capable, being flexible enough to support the
anonymization of most security relevant logs without major
modification; and (4) have a rich supply of anonymization algorithms
available for use on various fields. To meet these needs, we have
developed \flaim, the Framework for Log Anonymization and Information
Management.

\flaim strictly separates parsing from anonymization, providing an API
through which run-time dynamically loaded parsing modules can
communicate with the rest of the framework. Details such as file I/O
are abstracted away from \flaim's core, making it possible to handle streamed
data as easily as static data on the disk. The anonymization engine,
which consists of a suite of anonymization primitives for many
different data types, is also separated from the profile manager which
manages the XML based anonymization policies (e.g., parsing policies,
validating policies against schemas, etc). These three components,
together, provide an extensible and modular anonymization framework
able anonymize data to multiple levels for multiple types of logs.

While we have only talked about computer and network logs to this
point, the uses of a general framework for anonymization extend far
beyond the sharing of network logs. In late 2004, University of
California---Berkeley researchers were denied permission to analyze a
large set of personal data about participants in a state social
program after their systems were hacked and data on approximately 1.4
million people were breached \cite{poulsen04}. This example highlights
two points. First, research often depends on access to large amounts
of data. The UC-Berkeley team was investigating how to get better care
to homebound patients. Similarly, proponents of an NSA domestic spying
program have claimed that the 9/11 hijackers could be have been
identified by a program that analyzed communications data
\cite{gorman06}. Second, there are concerns of privacy protection on
these data sets seen in the backlash against such
programs. Anonymization potentially offers the best of both worlds,
allowing analysis while also protecting privacy. If the UC-Berkeley
data had been anonymized before being distributed, the vulnerability
to identity theft and other misuse may have been mitigated. While we
have not created modules for these other data sets yet in \flaim, we
have made \flaim's core fairly agnostic about the data source, capable
of working with any sort of record/field formatted data. This
generality opens up many possible future applications that we have not
even considered as of yet.

The rest of this paper is structured as follows. In
section~\ref{overview}, we present an overview of \flaim, discussing
its goals and functionality. We present a detailed description of the
\flaim architecture and API in section~\ref{arch}. In
section~\ref{anony}, we describe the anonymization algorithms
available in \flaim and provide an extended example of their use in
section~\ref{example}. Section~\ref{related} discusses other log
anonymization tools. We conclude and discuss future work in
section~\ref{conclusion}.

\section{Overview of \flaim and Goals}\label{overview}
There are four properties that an anonymization tool must have in
order to meet the varied needs of potential users. The anonymization
tool must: supply a large and diverse set of anonymization algorithms,
support many logs, support different levels of anonymization for a
log, and finally have an extensible, modular architecture.

These properties are not independent but to a large extent depend upon
each other. Namely, to support multi-level anonymization of logs it is
necessary to have multiple types of anonymization algorithms. In order
to provide support for multiple types of logs it is extremely useful
to have an extensible, modular architecture for the tool.

\subsection{Utility and Strength of Anonymization Primitives}
We use the term \emph{anonymization algorithm} or \emph{anonymization
primitive} to describe an algorithm that takes as input a piece of
data and modifies it so that it does not resemble its original form.
For instance, we could have an anonymization algorithm that takes
proper names, such as ``Alice'' and ``John'' and transforms each name
by constructing an anagram (e.g., ``John'' becomes ``Honj'' and ``Alice''
becomes ``Cliea'').

Clearly, anyone seeing these anonymized names would be able to guess
at the unanonymized names. Thus, we define the {\em strength} of an
anonymization algorithm to be related to the difficulty of retrieving
information about the original values from the anonymized values. For
instance, a black marker anonymization algorithm, which just replaces
every name with the string ``NULL'' is very strong---requiring one to use
only other identifying fields if they want to gather information about
the unanonymized value of that field.  In a similar way, we can talk
about the strength of a set of anonymization algorithms applied to
several distinct fields. For example, a scheme that only anonymizes
one field might not be strong enough, but one that anonymizes  a
set of 3 fields together may be much stronger.

Clearly, the strength of an anonymization algorithm is crucial to
making sure that private data is not retrieved from shared logs, but
this is not the sole goal of sharing logs. The real purpose to sharing
logs is to allow analysis of your logs. To do this, one must preserve
some type of information in the logs. The {\em utility} of an
anonymization algorithm is related to how much information is
preserved in the anonymized value. Note that just as we can talk about
the security of an anonymization primitive, we can also talk about the
utility of a scheme or set of algorithms applied to specific fields.

Consider the proper name anonymization example from above. The anagram
anonymizer has a greater utility than the black marker anonymization
primitive. With the anagram anonymized names, we lose the structure of
the word, but there still remains the number of letters and the actual
letters from the unanonymized value. The black marker anonymization
algorithm strips away all such information.

It is clear that the utility and strength of an anonymization
algorithm are strongly related. Furthermore, these relationships are
complex, since the utility and strength are based on the type of
analysis we are doing. In our future work we plan on further exploring
the relationship between the utility and strength of various
anonymization algorithms based on the task of detecting security
problems in network logs. However, a first step to reaching this goal
is creating a flexible tool like \flaim that allows us to make such
trade-offs in anonymization.

\subsection{Diverse Set of Anonymization Algorithms}
As we mentioned earlier, the central goal of sharing network and
system logs is to aid in the analysis of security related incidents on
the network. And essential to meeting this goal for various
organizations and their unique scenarios is the ability to make
trade-offs between the \emph{utility} of the anonymized log and the
\emph{strength} of the anonymization scheme. A necessary condition to
make these trade-offs is to have a diverse set of anonymization
primitives available for the tool at hand.

Earlier, we discussed how different anonymization primitives can make
different trade-offs between security and information loss with a
simple example. Now let us examine a more complex example with IP
addresses.  Figure~\ref{table:ipanony} shows a set of IP addresses
anonymized by various algorithms. Black marker anonymization maps all
the IP addresses to the same value, $c$. This leaves no scope for
analysis on that field. Random permutation maps each IP address to
another address at random. This allows some type of analysis, as we
could determine if two hosts in different records are the same.
Truncation, in this example, removes the last 16 bits of an IP
address. This allows one to see what the different domains are, but
several hosts are collapsed down into single values.  Thus, it becomes
difficult to separate individual hosts in the records.  Finally,
prefix-preserving anonymization maps the IP addresses to random
addresses, but it preserves the subnet structures. In this example, we
see that the $141$ class A network is consistently mapped to $12$,
preserving the subnet structures, but not revealing the original
subnets. In addition, it preserves individual hosts like a random
permutation, thus preserving much more utility than the other
methods. However, this comes at a cost of anonymization strength as we
have shown in \cite{Slagell05}.

Similar trade-offs can be made for other fields by having multiple
anonymization algorithms available.  Even the ability to decide which
fields are anonymized, provides a mechanism to make similar
trade-offs. However, having this vast set of anonymization primitives
is just one of the necessary conditions to create multiple levels of
anonymization, a point we discuss further in a later part of this
section.
\subsection{Supporting Multiple Logs}
A security incident is often only revealed in the presence multiple
logs.  Thus, proper analysis often requires access to many different
logs. Therefore, a holistic anonymization solution should anonymize
different types of logs. While one could write a separate tool for
each log type, this is a very inefficient approach.  This is
especially apparent when we notice that many logs share
the same set of common fields (e.g., IP address, port number,
timestamp). The only difference is often in how these fields are
represented within the logs (e.g., dotted decimal IP addresses in
netfilter logs and binary 32 bit unsigned integers in network byte
order in NetFlows).

\flaim supports multiple logs by having many ``modules''; each module
parses one specific type of log.  New modules can be created and added
to \flaim by merely implementing a simple interface that communicates
through the \flaim API. Modules are loaded as run-time libraries, so
no recompilation of \flaim is necessary. This allows users to leverage
the diverse set of anonymization algorithms that \flaim provides by
only creating a module to do the I/O and parsing.

\emph{Note: At this time \flaim modules exist for netfilter logs, pcap
  log support will be finished this summer, and a NetFlow module will
  be finished in the fall before USENIX Lisa.}
\subsection{Multi-level Anonymization}

Security related logs often need to be shared between several
different organizations to investigate a compromise, or even
internally between different groups within the organization. Logs
could also be shared externally with organizations providing security
out-sourcing services. The level of anonymization to be applied to the
logs changes with the recipient of the log. For internal users, the
logs may not have to be anonymized as strongly, whereas for external
users they most likely require stronger protections.

This means that an anonymization tool must support multiple levels of
anonymization. Different levels could be used for different recipients
as well as different situations for log sharing. Having multiple
anonymization algorithms of different strengths provides the building
blocks for multi-level anonymization, but something more is
needed. There should be a system to express and evaluate these
anonymization schemes.  \flaim does this through use of XML
anonymization policies. Users create these XML policies or use
predefined ones to choose the appropriate level of anonymization for a
given log. In particular, this policy will specify which fields are
anonymized and in what way.  Schemas built into \flaim check that the
policies are syntactically correct and that the options make semantic
sense for the type of fields. Since these policies are \emph{not} hard
coded into \flaim, they can be easily modified at any time.

\begin{table*}[tb]
\centering
\begin{tabular}{||l|c|c|c|c||} \hline
IP & Black Marker & Random Permutation & Truncation & Prefix Preserving \\ \hline
$141.142.96.167$ & $c$ & $141.142.132.37$ & $141.142.0.0$ & $12.131.102.67$\\
$141.142.96.18$ & $c$ & $141.142.96.167$ & $141.142.0.0$ & $12.131.102.197$\\
$141.142.132.37$ & $c$ &  $12.72.8.5$ & $141.142.0.0$ & $12.131.201.29$ \\
$12.161.3.3$ & $c$ & $212.3.4.1$ & $12.161.0.0$ & $187.192.32.51$\\
$12.72.8.5$ & $c$ & $141.142.96.18$ & $12.72.0.0$ & $187.78.201.97$\\
$212.3.4.1$ & $c$ & $12.161.3.3$ & $212.3.0.0$ & $31.197.3.82$\\ \hline
\end{tabular}
\caption{Example of 4 common methods to anonymize IP addresses}
\label{table:ipanony}
\end{table*}

\subsection{Modularity}
As we said, \flaim supports multiple log formats (e.g., netfilter,
pcap, etc). We could have done this with a large monolithic piece of
software code. In this case, one could use \flaim for many types of
logs, but still only a small set of all the logs out there, namely,
the set that we saw as important for our needs. However, we did not do
this because we wanted \flaim to be extensible.  We have, instead,
created a very modular framework with a strict API between the
separate components. Consequently, anonymization, policy verification
and parsing have all been separated. This allows not only us, the
\flaim developers, to add support for new types of logs, but it makes
it much easier for third party module developers to add support for
new types of logs within \flaim.

\section{FLAIM Architecture}\label{arch}

\begin{figure}[t] 
   \centering
   \includegraphics[width=3in]{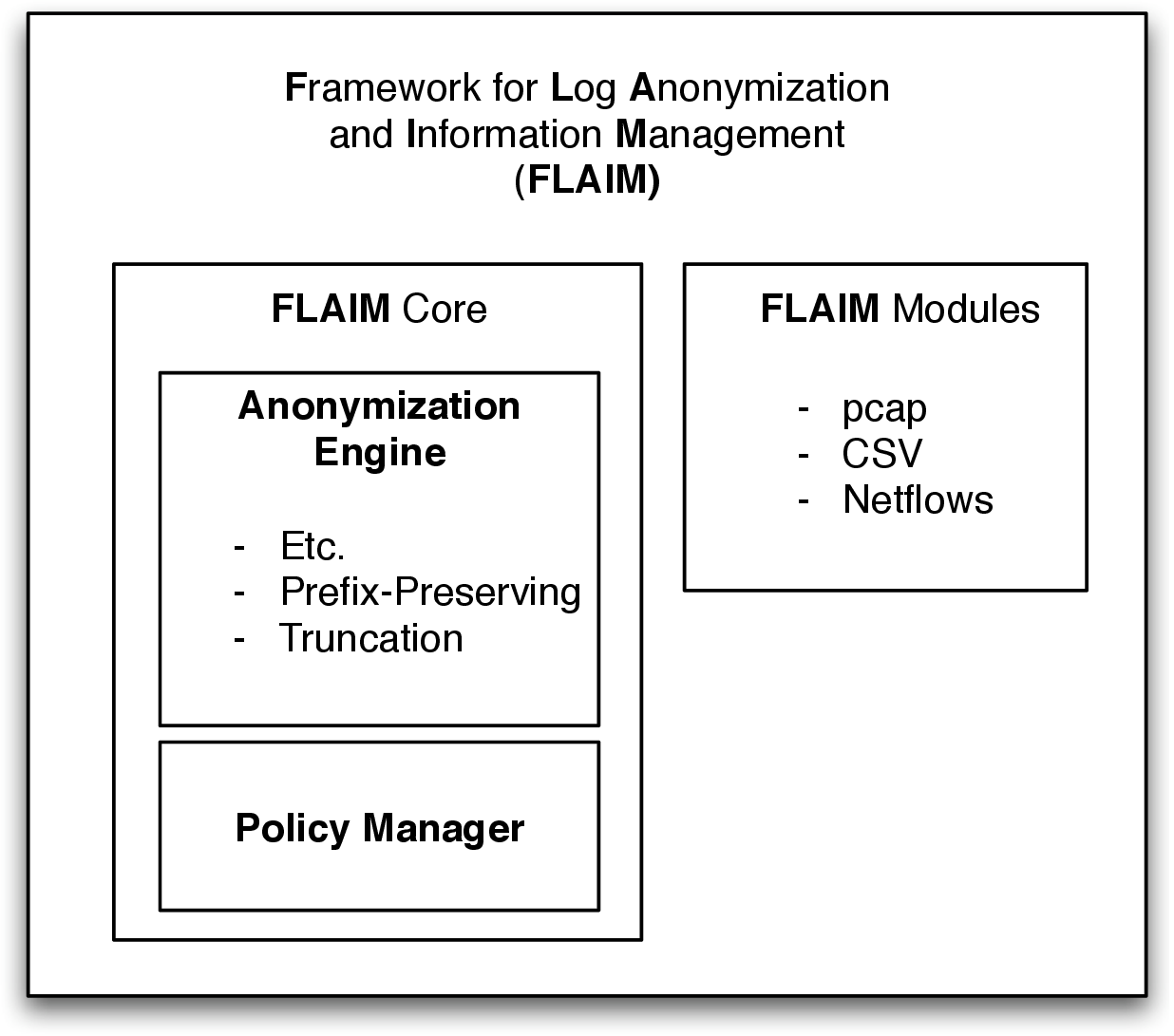} 
   \caption{The overall organization of FLAIM.}
   \label{fig:OverallOrg}
\end{figure}

To run \flaim, one must specify a \emph{user policy}---a description
of the anonymization algorithms to be used for each field. One of the
key contributions of \flaim is that it allows the user to specify an
anonymization policy that can make use of a multitude of anonymization
algorithms at run-time. Another key contribution of \flaim is that
parsing and I/O are separated to allow third parties add support for
additional logs. To achieve this, the architecture is composed of two
main components: \flaim Core and \flaim Modules. \flaim Core consists
of a suite of anonymization algorithms---the anonymization
engine---and the policy manager. \flaim Core reads and writes to
records via the Module API implemented by each I/O
module. Figure~\ref{fig:OverallOrg} illustrates the overall architecture of
\flaim.

\flaim modules consist of libraries of methods to parse various types
of logs. A single module is normally used to parse a single type of
log file. Every module is compiled as a dynamically linked library
which loads at runtime. Each module must implement the Module API
which defines a set of methods necessary for \flaim Core get records
to process and return them to the module to be reassembled and written
out.

In understanding the design of \flaim, it is useful to distinguish
between three different actors that play a role in the development and
use of \flaim. We have alluded to them before, but define them
formally below.

\begin{description}
\item\textbf{LAIM Group:} This refers to the Log Anonymization and
Information Management group at the NCSA. We developed \flaim, both
\flaim Core and the first \flaim Modules.
\item\textbf{Module Developers:} These people have developed or are
interested in developing log parsing modules that implement the Module API.
\item\textbf{\flaim Users:} These people are interested in using \flaim
to anonymize a set of logs using existing modules.
\end{description}

The basic workflow for \flaim is as follows:\\

\begin{enumerate}
\item \flaim is called with parameters that specify the input/output
data, the user policy, and the modules needed to parse the
input/output data.
\item The policy manager parses the user policy and determines the
anonymization algorithms that will be applied to each field of the
data. In addition to parsing, it uses schemas to validate the policy.
\item A record is ``requested'' by \flaim core via the Module API. 
\item The record is anonymized based on the user policy.
\item The anonymized record is sent back via the Module API and
  written out.
\item The last 3 steps are repeated for all records.
\end{enumerate}

In the rest of this section, we describe the Policy Manager and the
Module API. Also, at the heart of \flaim Core is the Anonymization
Engine which is basically a set of anonymization algorithms. The
anonymization engine was developed by the LAIM group and is being
continually extended. Currently, there are 8 different types of
anonymization algorithms implemented that differ based on their
strength. Section~\ref{anony} describes the various anonymization
algorithms in detail, and thus we defer further discussion of the
anonymization engine till the next section.

\subsection{Policy Manager}

The policy manager ensures that the anonymization policy specified by
the user is valid---that is, it specifies a known anonymization
algorithm with valid options for a data type that makes sense.  For
example, it would not allow prefix-preserving IP address anonymization
to be specified for a timestamp field. The anonymization policy must
be validated with both the FLAIM schema and the Module Schema to be
accepted.

The \flaim schema details what anonymization algorithms are available
as well as options for those algorithms. As new versions of \flaim are
released, new anonymization methods will be included in the
anonymization engine. The set of anonymization methods currently
available is listed in the \flaim schema. The \flaim schema is
maintained by the \flaim developers. 

The module schema indicates which anonymization methods are
appropriate for the specific fields in the log.  As the developers of
FLAIM, we can specify to what data types a particular anonymization
algorithm can be applied. However, we cannot anticipate the names or
handles that a module developer will use for those fields.  These
names are not only used in the anonymization policy, but also in the
meta-format that the module uses to send records to \flaim.  In
essence, \flaim Core simply sees a record as a list of tuples of the
form $<\mathit{FieldName},\mathit{FieldValue}>$. \flaim core then
matches the field name with the options specified for the field name
in the policy to determine how to anonymize it.

We could construct a list of valid field names from which the module
developer could choose names. This would allow \flaim Core to
recognize the data type and make sure the algorithm being applied
makes semantic sense for that field. However, we could hardly
anticipate the needs for any type of log.  For example, we could
specify ``IP1'' and ``IP2'' as valid names, however, a log that we do
not anticipate may have four distinct IP addresses per record.  It
could also have fields of types we didn't expect, but that need to be
passed to the anonymization engine to be kept with the records if they
are being reordered.

Consequently, we do not restrict the set of field names that the
parser could use when sending records, i.e the same field names
specified in the user's policy.  So to make sure that an algorithm is
applied to the correct data type, the module developer creates a
schema using a very simple syntax that specifies what algorithms can
be used with what field names or handles.  Being the only one who
knows the data type that corresponds to the field names, this must be
done by the module developer. We will distribute header files for the
anonymization algorithms so that module developers know the data type
expected by an anonymization algorithm.

\begin{figure*}[t] 
  \centering
  \includegraphics[width=4in]{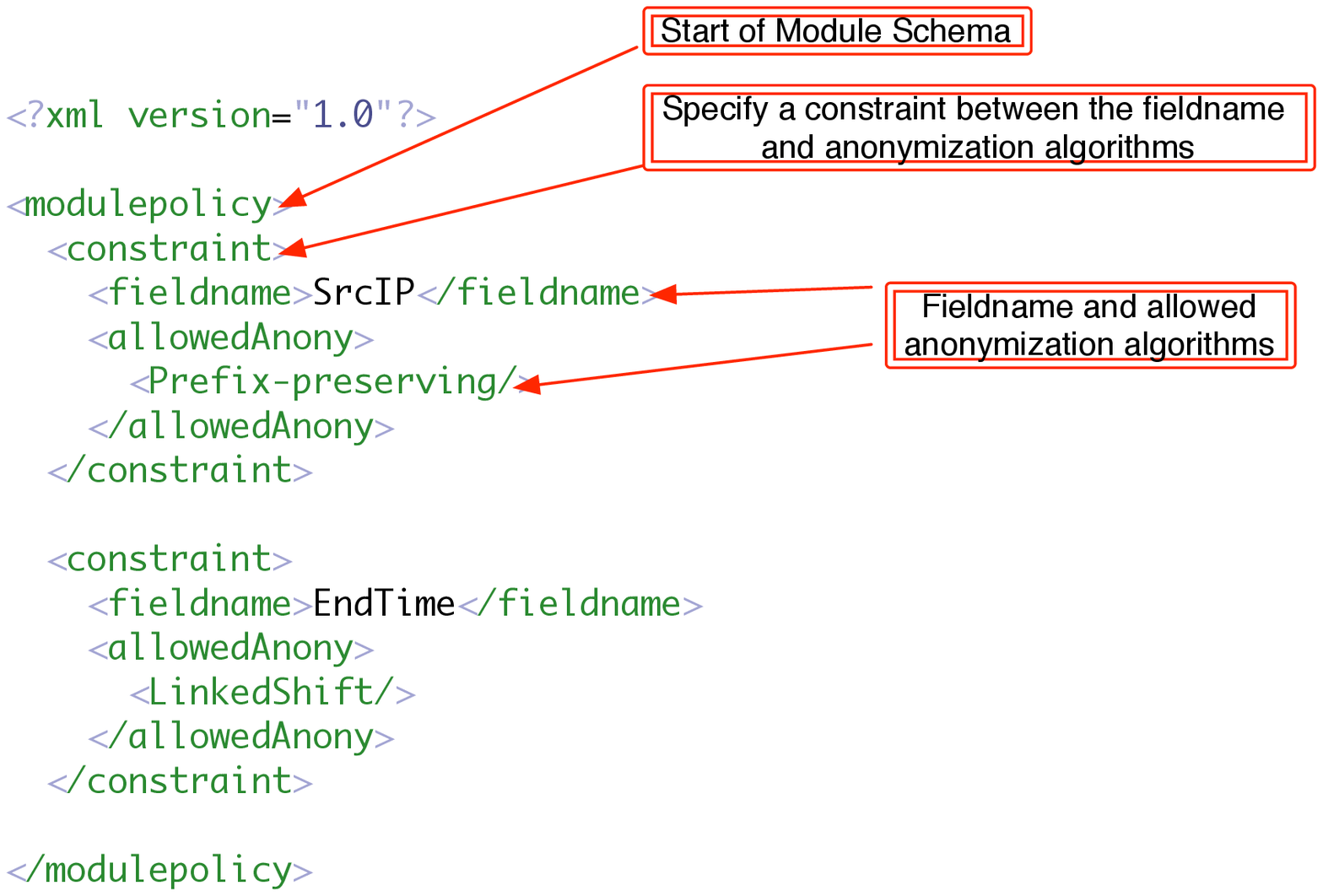} 
  \caption{A Module schema written in the \emph{Module Schema Language}}
  \label{fig:MSinMSL}
\end{figure*}

The module schema is written in XML and must conform to our definition
of a ``Module Schema Language''. For module developers who have
experience with Schematron, the module schema may be written in
Schematron. We use a XSLT stylesheet to translate the Module schemas
written in a simple Module Schema Language to Schematron
stylesheets. Figure~\ref{fig:MSinMSL} is an example Module Schema
Policy written in the Module Schema Language.

\begin{figure*}[t]
   \centering
   \includegraphics[width=4.5in]{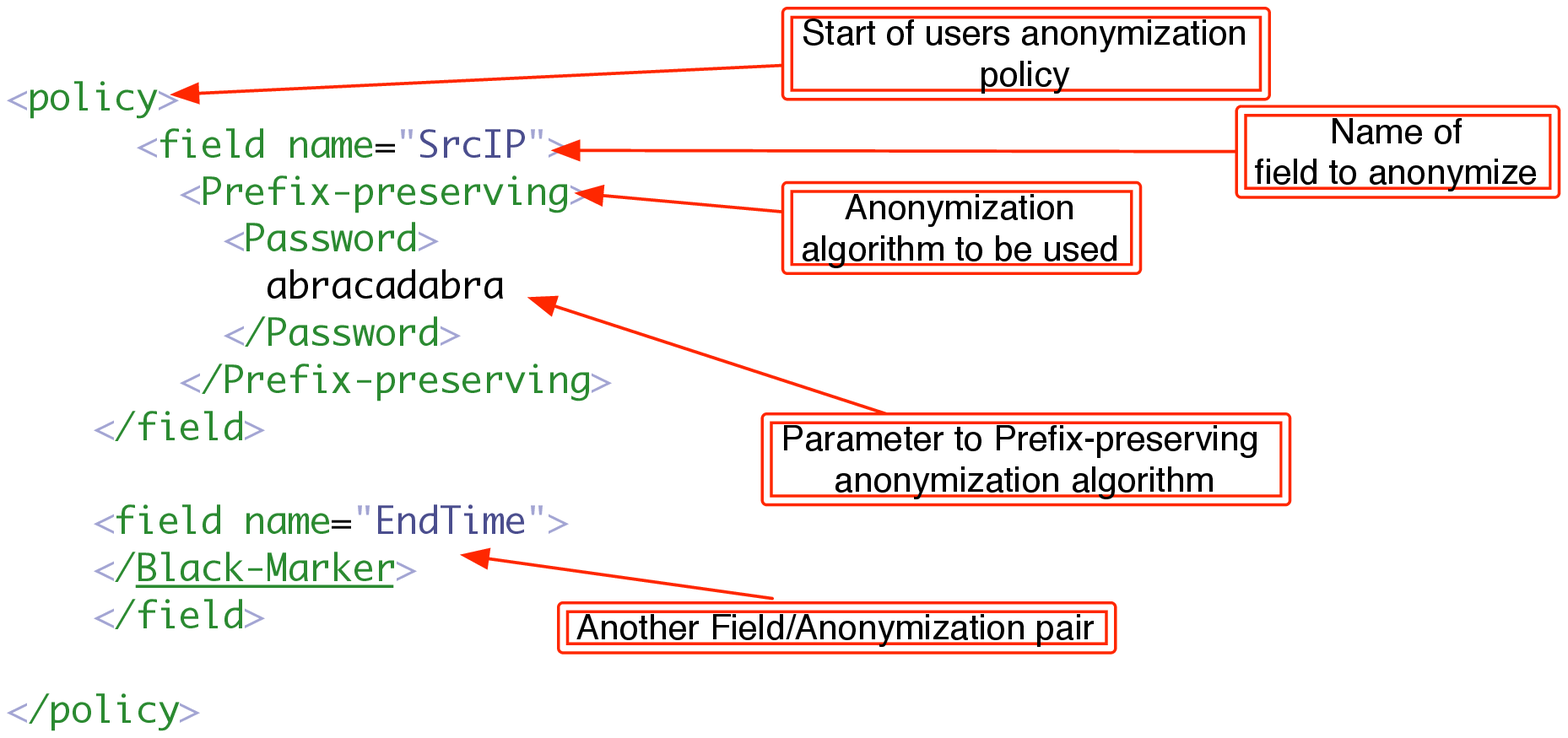} 
   \caption{A simple user anonymization policy. }
   \label{fig:simpleUser}
\end{figure*}

We see that, together, validation against both the \flaim schema and
the module schema ensures that the user's anonymization policy uses
anonymization algorithms that are supported by \flaim, with
appropriate options, and are applied to fields in a semantically sound
manner (e.g., so an algorithm that only makes sense for time stamps
is not applied to IP addresses).
\\
Thus, the policy manager's role is two fold:
\begin{enumerate}
\item Validate the user's  policy against the \flaim schema and the
  module schema. 
\item Parse the user policy file into an internal format to be used by
  the anonymization engine.
\end{enumerate}

Figure~\ref{fig:simpleUser} is an example of part of a sample user policy.

\subsection{I/O Modules}
The I/O library reads and parses an input file or stream.  A module
library's purpose is to abstract away from \flaim the task of
physically manipulating the medium which contains the data. All
modules must implement the module API that is defined by the LAIM
developers. The API allows \flaim to access the module schema,
described above, as well as provides methods to access the
data. Modules are responsible for handling storage, retrieval and
parsing of the data. Of these data, there are two important types: static files
or streams.

Static files are ordinary files to which \flaim has random
access---the key distinction from streams being random
access. However, it is not always practical to store all the data in
static files. Rather, real-time anonymization may be done on the data
as it is collected, perhaps even before it is written to disk. Data
that can only be accessed once is called ``streaming data'', a
particular data source being called a ``stream''.

Handling a stream can be quite different from a static file. For
example, an anonymization algorithm may require scanning the log twice
and hence the ability to go back to the beginning of the data source.
This can be problematic for streams, of course. As such, there are
mechanisms in the \flaim API to let the module know if one of the
anonymization algorithms chosen will need random access. It is up to
the module developer to decide whether or not to support such
algorithms and how to do so.

\subsection{The Module API}
The module interface has been designed to be as simple as possible. It
consists of five functions that a module must implement. These
functions control the input and output of records as well as resetting
the counter in the file. Procedures that need to be implemented are
listed below.  \lstset{language=C, basicstyle=\small}

\begin{lstlisting}
// Returns the name of the file containing the module schema
char* getModuleSchema()

// Communicate to the module the command line parameters passed into FLAIM
void setDataSets(char* inputfilename, char* outputfilename)

// Return a single record
Record getRecord()

// Get the current location of the record counter.
int getCntrValue()

// reset the record counter
bool resetCntrValue();

// return true if this is the last record
bool atEnd();

// Write a single record
int putRecord(Record r)
\end{lstlisting}

\subsection{Summary of the Architecture}
The architecture of \flaim is designed to satisfy the properties
stated in the previous section. The Anonymization Engine in \flaim
CORE contains a diverse set of anonymization algorithms for many
different data types. The Policy Manager component of \flaim Core
allows users to easily provide many different levels of anonymization
to logs by utilizing the many algorithms in the Anonymization
Engine. Users can easily change the anonymization of a log by
specifying a new user policy in the simple XML language specified by
\flaim.  Users can create anonymization polices tailored to the log
and situation in which the log is being distributed.

Finally, \flaim is designed in a modular fashion to be highly
extensible to many types of logs. Log parsing modules can be
dynamically loaded into \flaim at runtime. This allows new logs to be
added without changing the anonymization engine or the policy manager.
We defined a simple API that all log parsing modules must
implement. \flaim Core will interact with the log parsing module via
this API. The API is simple, giving module developers much
flexibility. Taken as a whole, these components and the modules we have
created make \flaim the first extensible, multi-level, multi-log
anonymization tool with the largest set of anonymization primitives of
which we are aware.

\section{Anonymization Algorithms for \flaim}\label{anony}
\flaim implements many types of anonymization algorithms. A few are
coupled very tightly with the data type being anonymized, but many can
be applied to multiple fields. However, the default values and other
configuration options may be affected by the data type. For example,
while truncation can be applied to almost any field, it makes sense to
truncate a MAC address by 40 bits but not to do the same to a 32 bit
IPv4 address. Below we discuss the different classes of anonymization
algorithms \flaim supports, and in the following section we go into an
extensive example of how they are used.

\subsection{Black Marker}
{\em Black marker} anonymization---a term we coined in
\cite{Slagell04b}---is equivalent, from an information theoretic point
of view, to printing out a log and going over each value of a
sensitive field with a black marker. This analog variant is often seen
in sensitive documents retrieved from the government. We simply
implement a digital equivalent. This could mean that we just delete
the field.  However, that could break log analysis tools for
anonymized logs.  So, when performing black marker anonymization on a
field, we replace all values with a single canonical value for that
field. For example, all IP addresses may be replaced with ``0.0.0.0''.

\subsection{Truncation} 
{\em Truncation} works by taking a field and selecting a point after
which all bits are annihilated.  For a string value, you choose some
middle point---not necessarily defined as a fixed number of characters
from the beginning---and cut the string off after that point. For
example, one could truncate the domain information from e-mail address
so that ``user@example.net'' is replaced with simply ``user''. For
binary values, a fixed point is chosen from the beginning of the
value, and all bits after that point are set to 0---keeping the binary
value the same length.  So truncating after 24 bits on a binary
representation of an IP address gives the class C subnet. Note that
truncation of all bits simply result in a specific case of black
marker anonymization. Consequently, these algorithms are not mutually
exclusive.

\subsection{Permutation}
In the most general sense, a {\em permutation} is a one-to-one and
onto mapping on a set. Thus, even a block cipher is a type of
permutation. There are many ways that permutations can be used. For
larger binary fields, it usually makes sense to use a strong,
cryptographic block cipher. Thus, if one wishes to use the same
mapping later, they just save the key.  This is excellent for fields
like the 128 bit IPv6 address.  However, there are no strong 32 bit
block ciphers to do the same for IPv4 addresses.  Using a larger block
would require padding, and the output of the anonymization function
would be larger than the input. In these cases, one can use tables to
create random permutations. The problem is, of course, that one cannot
save these tables as easily as a cryptographic key to keep mappings
consistent between logs anonymized at different times.

In addition to random permutations that are cryptographically strong,
there is sometimes a need to use structure preserving permutations for
certain types of data. When differences between values must be
preserved---often the case with timestamps---a simple shift can be
used. Shifting all values by a certain number may be an acceptable way
to permute values in some instances. For IP addresses, the subnet
structure may need to be preserved without knowing the actual subnets
(e.g., when analyzing router data). Prefix-preserving pseudonymization
is appropriate here. This is a type of permutation uniquely applied to
IP addresses, and we discuss it in detail in an example of anonymizing
netfilter logs in the next section.

\subsection{Hash} 
Cryptographic {\em hash} functions can be useful for anonymization of both
text and binary data.  The problem with binary data, of course, is
that one must often truncate the result of a hash function to the
shorter length of the field. This avoids breaking log analysis tools
that operate on those logs. This also means that the hash function is
weaker and has more collisions.  For fields 32 bits or smaller,
dictionary attacks on hashes become very practical. For example, it is
well with the capability of a modern adversary to create a table of
hashes for every possible IPv4 address. The space of possible values
being hashed is too small. Thus, hash functions must be used carefully
and only when the possibility of collision in the mappings is
acceptable. Often, it is better to use a random permutation. For
string values, \flaim outputs the hash in an ASCII representation of
the hexadecimal value.

\subsection{HMAC}
{\em HMACs} (Hash Message Authentication Codes) are essentially hashes
seeded with some secret data. Adversaries cannot compute the HMAC
themselves without access to the key, and thus they cannot perform the
dictionary attacks that they could on simple hashes. In cases, where
hashes are vulnerable to such attacks, HMACs make sense.  However,
HMACs, like hashes, are inappropriate when collisions are
unacceptable.

\subsection{Partitioning} 
{\it Partitioning} is just what it sounds like.  The set of possible
values is partitioned into subsets---possibly by a well-defined
equivalence relation---and a canonical example for each subset is
chosen. Then, the anonymization function replaces every value with the
canonical value from the subset to which it belongs. Black marker
anonymization---how we implement it---is really just a special case
where there is only one subset in the partition, namely, the entire
set. Even truncation becomes a type of partitioning. For example, say
that the last 8 bits of IPv4 addresses are truncated off.  Then the
set of IP addresses is being partitioned into class C networks.
Furthermore, the canonical representation is simply the network
address of the class C subnet. However, partitioning is not always so
simplistic, and our next type of anonymization algorithm is a very
unique type of partitioning for timestamps.

\subsection{Time Unit Annihilation} 
{\em Time unit annihilation} is a special type of partitioning for
time and date information.  Timestamps can be broken down into year,
month, day, hour, minute and second subfields. When this is done, one
can annihilate any subset of these time units by replacing them with
0. For instance, if she annihilates the hour, minute and second
information, the time has been removed but the date information
retained---actually, a type of truncation.  If she wipes out the year,
month and day, the date information is removed but the time is
unaffected. If one annihilates every subfield, she ends up performing
a type of black marker anonymization on the timestamps. It is clear
that this is a very general type of partitioning, but it still cannot
partition in arbitrary ways. For instance, it cannot break time up
into 10 minute units.

\subsection{Enumeration} 
{\em Enumeration} can be very general, though \flaim currently uses it
as just an option for timestamps.  However, enumeration would work on
any well-ordered set. Enumeration, will first sort the records based
on this field, choose a value for a first record, and for each
successive record, it will choose a greater value. This preserves the
order but removes any specific information.  When applied to
timestamps, it preserves the sequence of events, but it removes
information about when they started or how far apart two events are
temporally.  A straightforward implementation could sort, choose a
random starting time, and space all distinct timestamps apart by 1
second.

\section{Netfilter Anonymization: An Extended Example}\label{example}
In this section, we look at how we have applied the various
anonymization algorithms described above to netfilter logs---the first
type of log we supported with \flaim. Since all of these fields are
also found in pcap traces, we can apply these same algorithms to pcap
logs ({\it support for pcap should be finished in the summer of
2006}). We describe which algorithms make sense for each field and how
they are uniquely adapted to each data type (e.g., how default values
change). {\it Note to reviewers.  This section can be read later or
skimmed without interruption to the rest of the paper.}

\subsection{Time Stamps}
\flaim supports 3 methods of timestamp anonymization. As with any data
type, we must choose a canonical form for the field.  We are using the
traditional UNIX epoch format, i.e. the number of seconds since Jan
$1^{st}$, 1970. This is fastest for 2 of the 3 applicable anonymization
algorithms.  Because timestamps often occur in pairs, with a starting
and ending time, we optionally allow a handle to a secondary time
stamp field to be specified in the anonymization policy.  If the
secondary timestamp is specified in the anonymization options, it is
adjusted with the first timestamp so that the difference between the
two values remains the same. For example, this could be used to keep
flow duration the same in NetFlows.

\subsubsection{Time Unit Annihilation}
\flaim can use time unit annihilation, as described earlier in this
section, to anonymize timestamps. Timestamps are converted in the
anonymization engine from the canonical format to an internal
structure to perform this type of anonymization. Then they are
converted back.

\subsubsection{Random Shift}
In some situations it may be important to know how far apart two
events are temporally without knowing exactly when they happened.  For
this reason, a log or set of logs can be anonymized at once such that
all timestamps are shifted by the same random number, in seconds.  We
call this method of anonymization a {\it Random Time Shift}. As noted
in the previous section, this is a special type of permutation.

We allow a lower and upper bound for the random number to be set for
two reasons. First, an upper bound can prevent a shift so far into the
future that it overflows the 32 bit timestamp field and events
wrap-around back to Jan. $1^{st}$, 1970. Second, by setting the lower
and upper bound equal, you can control exactly how much the time
stamps shift. This allows you to keep the anonymization mapping
consistent between different runs of \flaim.

\subsubsection{Enumeration}
We implement the enumeration method of anonymization described in
general terms earlier in this section. \flaim chooses a random
starting time for the first record, and each subsequent timestamp---if
differing from the previous---is 1 second later. The end result is
that one can tell if two records happened at the same time or one
before the other, but they can know nothing else. Implementing the
enumeration algorithm exactly as described does pose one problem,
however.  The presorting can be slow on even medium sized logs and
nearly impossible with streamed data. Though logs are not always
presorted, they are often close to being in order. Often they are
simply out of order because of clock skew between different data
sources.

In our implementation, we exploit the fact that timestamps are often
just slightly out of order (e.g., NetFlows are nearly in order by
ending timestamp) and buffer events to sort locally.  This buffer is
used like a sliding window in which only events within the window can
be sorted.  Events before the window are written out already, and
events after window have not yet been read. Since events are usually
not terribly disordered, this sorts records with great accuracy often
when using a small window.  The size of the window is user-selectable.
Larger windows must be used for logs that are more disordered. We have
found this to be a useful way to allow users to select a level of
compromise between efficiency and accuracy.

\subsection{IP Addresses}
We have implemented four types of IP address anonymization in \flaim:
truncation, black marker, prefix-preserv-ing, and random
permutation. Nearly all of the log anonymization tools out there use
one of these algorithms or hashing---which can be easily brute-forced
on the small 32 bit space for IPv4 addresses. Others are slight
variations, such as permutations that fix a hard-coded set of internal
IP addresses. In FLAIM, the canonical form for an IP address is a 32
bit unsigned integer.

\subsubsection{Truncation and Black Marker} \flaim will truncate IP
addresses from 1 to 32 least significant bits. Truncated bits are
replaced with 0's. Thus truncating all 32 bits results in black marker
anonymization with every IP address being replaced by
``0.0.0.0''. While the common choices would be to truncate on an 8 bit
boundary, \flaim will truncate anywhere between 1 and 32 bits.

\subsubsection{Random Permutation} We also support anonymization by
creating random permutations on the set of possible IP addresses. This
permutation is then applied to each IP address in the log. We
implement this algorithm through use of two hash tables for efficient
lookup. One hash table is used to store mappings from unanonymized to
anonymized IP addresses.  The other hash table is used to store all of
the anonymized addresses so we can check if an address has already
been used when creating a new mapping.  Because the first table is
indexed by unanonymized addresses, the whole table would have to be
searched for a free anonymized address if we did not use a second hash
table.  In this way, we trade a little storage for a large
computational speed-up. For even higher space efficiency we could use
a Bloom filter \cite{Bloom70} to store the set of used IP addresses.
Since Bloom filters never give a false negative, we would not map two
distinct IP addresses to the same value, and thus our function would
remain injective.

Sometimes it is desirable to fix certain elements within the
permutation.  Say that an organization wants all external IP addresses
to remain unchanged.  This sort of less random permutation can be
implemented by simply pre-filling the tables with entries that fix
this subset of elements.  Future versions of \flaim may allow one to
specify a CIDR addressed subnet to fix in the permutation.

\subsubsection{Prefix-preserving Permutation} Prefix-preserving
pseudonymization uses a special type of permutation that has a unique
structure preserving property.  The property is that two anonymized IP
addresses match on a prefix of $n$ bits if and only if the
unanonymized addresses match on $n$ bits. This preserves subnet
structures and is often preferred to random permutations, but the
simple fact that there are many times fewer permutations of this type
makes it weaker.  As a general principle, cryptographic algorithms
that preserve structure are more open to attack, and this algorithm is
not an exception.  For example, an adversary that injects traffic to
be recognized later not only gleans information about the addresses
she specifically attacked, but she also learns many of the
unanonymized bits of addresses that share prefixes with the addresses
she attacks. For this type of anonymization we implemented the
Crypto-PAn \cite{Xu01,Xu02} algorithm and generate keys by hashing a
passphrase the user provides. In this way, tables are not used, and
logs can more easily be anonymized in parallel across different
locations. This is a sensible choice when an injection attack is
unlikely to be effective (e.g., a one-time release of logs with a
particular key) and the subnet structure is very important.

\subsection{MAC Addresses}
MAC addresses have traditionally been globally unique, and the first 3
of 6 bytes are usually indicative of the network card manufacturer. As
such they are somewhat sensitive. However, proliferation of
virtualization means that MAC addresses are not always globally unique
now. In addition, many hardware devices allow you to change the MAC
(e.g., SOHO routers). \flaim supports 3 types of anonymization for MAC
addresses and uses a 6 byte unsigned char array as the canonical
representation.

\subsubsection{Truncation and Black Marker}
 \flaim will truncate any number of least significant bits replacing
them with 0's. This could allow one to remove all identifying
information but the manufacturer or allow one to black marker
anonymize the whole address. While truncating 24 or 48 bits seems the
most useful choices, \flaim will truncate any number from 1 to 48
bits.

\subsubsection{Random Permutation}
It may be important to distinguish network interfaces within a log,
but not to know the specific MAC.  In fact, this is often the case
since knowing the specific MAC usually does not get you any
information except for the manufacturer unless you have access to special
outside knowledge (e.g., access to ARP or DHCP logs). For this reason
we have an algorithm that creates a random permutation of MAC
addresses. It is implemented in the same manner as the random
permutation of IP addresses.  In fact, this function is just a wrapper
for a more general function---a practice we make use of whenever possible.

\subsection{Hostnames}
\flaim can also anonymize hostnames that are both local and fully
qualified with domain names. The canonical form of a hostname is a
string. If there are periods in the hostname, the host part is to the
left of the first period and the domain name is the part to the right
of the first period. Whether or not the hostname is fully qualified,
can make a difference in the anonymization function.

\subsubsection{Black Marker}
Black marker anonymization replaces fields with constants. In the
context of hostnames, \flaim can be configured to black marker
anonymize just the host part or the entire name. If it is configured
to black marker anonymize just the host part, the host part of the
name is replaced with the string ``host''. If the hostname is fully
qualified, the domain name is left untouched.  For example, a hostname
of ``vorlon'' would be replaced with ``host''. A hostname of
``vorlon.ncsa.uiuc.edu'' would be replaced with
``host.ncsa.uiuc.edu''.

Black marker anonymization can also be configured to replace the
entire hostname. If the hostname is fully qualified, it is replaced
with ``host.network.net'', otherwise it is simply replaced with
``host''. In addition to setting the black marker anonymization to the
host or full name, one can specify the constant strings with which to
replace the names.

\subsubsection{Hash}
Another anonymization algorithm available for hostnames is a simple
hash converted into ASCII output.  While we could hash just the host
part and leave the network part alone, this is not very useful since
the valid hostnames on a given network can be easily enumerated in
most cases, and thus the hash function could be brute-forced by a
simple dictionary attack. Therefore, we only hash the entire string.

\subsubsection{HMACs}
As noted earlier in this section, HMACs can be used instead of hashes
when there is concern of brute-force attacks.  Future revisions of
\flaim will add support for using HMACs to anonymize hostnames.

\subsection{Port Numbers}
We have implemented three methods of anonymization for port numbers in
\flaim.  The canonical representation for port numbers is a 16 bit
unsigned integer.

\subsubsection{Black Marker}
\flaim supports black marker anonymization of this field, replacing
every port number with port 0.

\subsubsection{Bilateral Classification} 
We call the second method of port number anonymization we implement
{\it bilateral classification}.  This is really just a special type of
partitioning. Often the port number is useless unless one knows the
exact port number to correlate with a service.  However, there is one
important piece of information that does not require one to know the
actual port number: whether or not the port is ephemeral.  In this way
ports can be classified as being below 1024 or greater than or equal
to 1024.  The canonical value for privileged ports is 0, and the
canonical value for ephemeral ports is 65535.

\subsubsection{Random Permutation}
In some cases it may not be necessary to know what the port is, but
only that some ports are seeing particular traffic patterns.  Such is
the case in detecting worms and P2P traffic.  Such traffic has unique
characteristics, and knowing the exact port number may not not even be
that useful. For example, a new worm may appear on an unexpected port.
If you are just counting on the port number, you will not recognize
this new worm.  However, if you look for worm-like behaviors appearing
on a fixed port, you can detect the worm \cite{Yin05}.  Similarly,
knowing the specific port of a piece of P2P software is becoming less
useful, as the ports are becoming dynamic.  On a particular machine
the port will be fixed for a while, but that particular machine often
chooses a random port. Thus, detection must depend on elements other
than knowing the specific port number.

Randomly permuting the port numbers does not affect the ability to
detect traffic worm or P2P traffic using these more advanced
methodologies since such services do not always use the same
predictable port numbers.  However, bilateral classification or black
marker anonymization may affect such methodologies as it is no longer
easy to separate the traffic from different applications on the same
machine.  For example, a machine may have a P2P application on port
7777 and a web server on port 8080.  Depending on the log and the
specific behavior of the P2P application, it may be difficult to
detect the P2P application as all of its traffic is aggregated with
the legitimate web server.  In this situation, a random permutation
provides the maximum amount of anonymization possible to complete the
analysis.

\subsection{Network Protocol}
\flaim supports black marker anonymization of the protocol field.  The
canonical form of this field is is an unsigned 8 bit integer
corresponding to the values assigned by IANA \cite{IANA} (e.g., TCP is
6, UDP is 17, ICMP is 1). If this field is anonymized, all protocols
are replaced with 255, the IANA reserved protocol number. For many
network logs it makes no sense to anonymize this field alone.  For
example, in a pcap log the TCP headers will give away the protocol
even if the protocol field is eliminated in the IP header.

\subsection{IP ID Number}
We support black marker anonymization of ID numbers because their use
is moderately sensitive to passive OS fingerprinting. All ID's are
replaced with 0 in this case. The canonical form of this field is an
unsigned 16 bit integer.

\subsection{IP Options}
We support black marker anonymization of IP options because their use
is moderately sensitive to passive OS fingerprinting and because they
can carry significant information in covert channels. Specifically, we
can see their usefulness in data injection and probing attacks. The
canonical form for this field is a string in the same format that
iptables uses to represent this variable length field. If black marker
anonymization is chosen, \flaim replaces the string with the null
string.

\subsection{Misc IP Fields}
We support black marker anonymization of Type-of-Service and
Time-to-Live IP fields because their use is very sensitive to passive OS
fingerprinting. Their canonical form is the same as the protocol
field, and we replace all values with 255 when performing this type of
anonymization.

\subsection{Don't Fragment Bit}
We support black marker anonymization of Don't-Fragment bits because
this field is very sensitive to passive OS fingerprinting. Nothing
smaller than a byte is sent to FLAIM, so any flags or boolean values
have the canonical form of an unsigned char.  If the value is
non-zero, it is treated as if the bit is set to true.  If the value is
zero, it is interpreted that bit is not set or set to false.  Thus,
when \flaim anonymizes this field, it simple replaces instances of the
field with 0.

\subsection{TCP Window Size}
We support black marker anonymization of the TCP window size because
its use is very susceptible to passive OS fingerprinting. The
canonical form of this fields is a 16 bit unsigned integer.  If
anonymization of this field is done, all window sizes are set to 0.

\subsection{Initial TCP Sequence Number}
The canonical form of this field is an unsigned 16 bit
integer. Because initial TCP sequence numbers are moderately sensitive
to passive OS fingerprinting, \flaim can be used to black marker
anonymize TCP sequence numbers, replacing all sequence numbers with 0.

While doing this is not syntactically breaking logs, it will lead to
semantically nonsensical values. Log analyzers should not break {\em
while parsing} logs anonymized in this way, but their output could be
quite unpredictable when fields like this are anonymized in such a
manner. More complex anonymization algorithms that try to identify
when packets are part of the same flow could be used, instead. The
draw back is that the anonymization of one field becomes dependent on
other fields within a record. \flaim is certainly general enough to do
this, and we do in fact handle relations between timestamp fields
within a record. However, for a field that has seen lesser demand for
anonymization, we have chosen to perform a simpler type of
anonymization in this first instance of FLAIM. User demand will
determine if more complex solutions are desired for some of these
fields.

\subsection{TCP Options}
We support black marker anonymization of TCP options because their use is
moderately sensitive to passive OS fingerprinting and because they can
carry significant information in covert channels. Specifically, we can
see their usefulness in data injection and probing attacks. The
canonical form for this field is a string in the same format that
iptables uses to represent this variable length field. If black marker
anonymization is chosen, \flaim replaces the string with the null string.

\subsection{ICMP Codes and Types}
NetFlow records and firewall logs can indicate both the ICMP code and
type without the packet data.  Truncation makes no sense since there
is no structure to the specific code and type numbers. Permutations
and hashing seem to offer no utility over simple black marker
anonymization. Thus, \flaim currently supports just black marker
anonymization of these fields. The canonical form for these fields is
the unsigned char. The value with which these fields are replaced is
0.

\begin{figure*}[t] 
   \centering
   \includegraphics[width=4in]{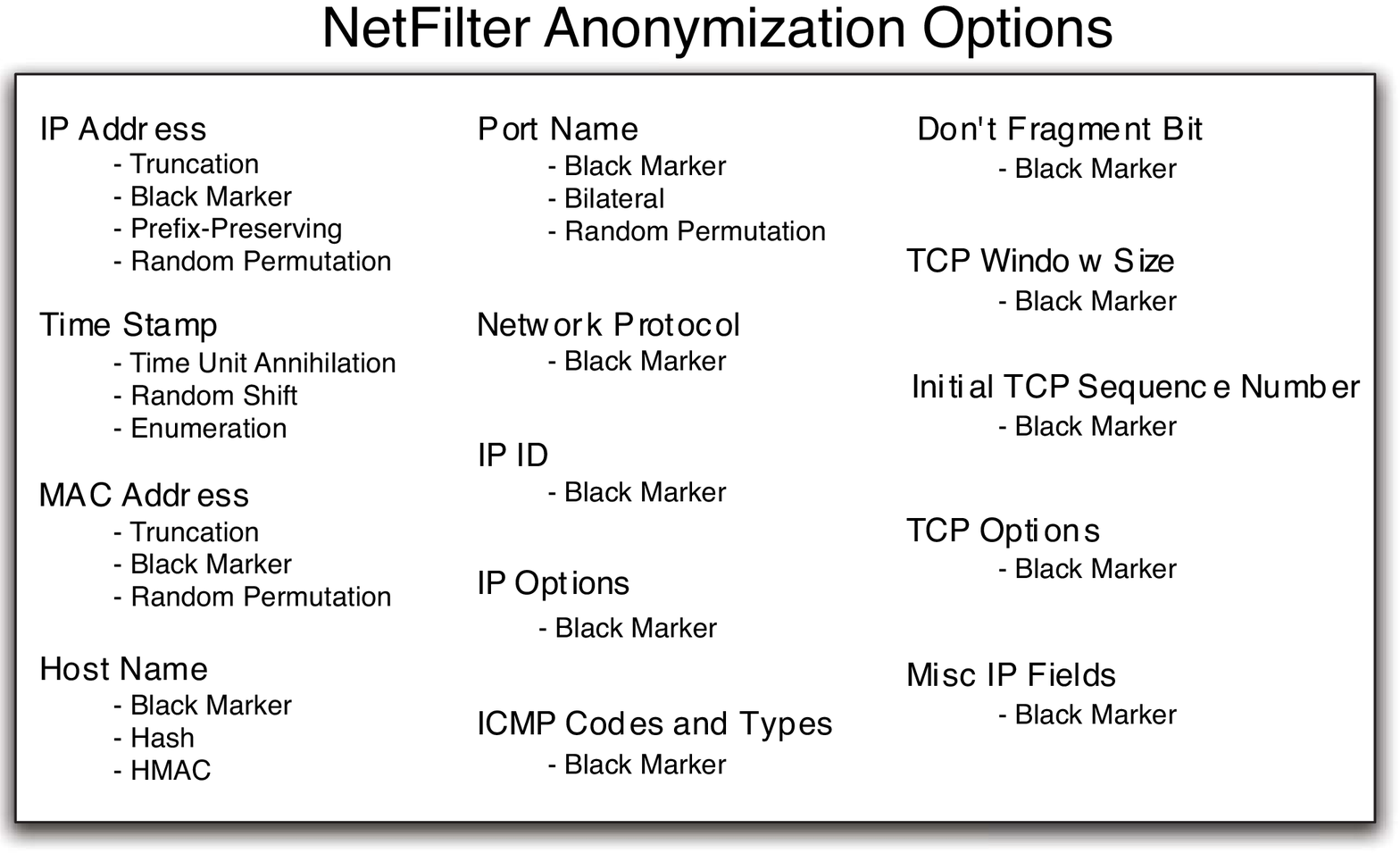} 
   \caption{Listed are the all the available anonymization options for the fields in Netfilter logs. A user can specify 324 different anonymization policies based on the anonymization algorithms listed above.}
   \label{table:iptableoptions}
\end{figure*}

Figure~\ref{table:iptableoptions} summarizes the anonymization methods that can be applied to the fields in the IP table logs.

\section{Related Work}\label{related}

While there have been several anonymization tools created for specific
logs, they have not been very flexible to date. The anonymization
primitives used have mostly been simplistic, and anonymization is
typically done on only one field with no options as to what
anonymization algorithm is used. Thus, in the current state of
matters, we have a collection of ad hoc tools created for the specific
needs of individual organizations, rather than flexible tools used by
many.

One of the major results in log anonymization---one that changed how a
lot of tools anonymize data---was the development of prefix-preserving
IP address anonymization.  The most commonly anonymized field in
security and network relevant logs is the IP address. It was long
desired to both preserve the subnet structure like truncation, and
also have a one-to-one mapping between anonymized and unanonymized
addresses. The solution was to create prefix-preserving permutations
on IP addresses.  Mathematically, we define such a mapping as
follows. Let $\tau$ be a permutation on the set of IP addresses, and
let $P_n()$ be the function that truncates an IP address to $n$
bits. Then $\tau$ is a {\it prefix-preserving} permutation of IP
addresses if $\forall$ $1\leq n\leq 32$, $P_n(x)=P_n(y)$ if and only
if $P_n(\tau(x))=P_n(\tau(y))$.

In recent years, several tools have been made that make use of this
newer, structure preserving form of IP address anonymization. Many are
based on tcpdpriv, a free program that performs prefix-preserving
pcap trace anonymization using tables. Because of the use of
tables, it is difficult to process logs in parallel with this tool. In
\cite{Xu01,Xu02}, Xu et al.\ created a prefix-preserving IP
pseudonymizer that overcomes this limitation by eliminating the need
for centralized tables to be shared and edited by multiple
entities. Instead, with their algorithm Crypto-PAn, one only needs to
distribute a short key between entities that wish to pseudonymize
consistently with each other. Their work made prefix-preserving
pseudonymization much more practical. In \cite{Slagell04}, we used
Crypto-PAn with our own key generator to perform prefix-preserving IP
address pseudonymization on a particular format of NetFlow logs. We
later implemented Crypto-PAn in Java as part of more advanced NetFlow
anonymizer we call CANINE \cite{Slagell05b}. CANINE supports several
NetFlow formats and anonymizes the 8 most common fields within
NetFlows.

In \cite{Sobirey97}, Sobirey et al.\ first suggested privacy-enhanced
intrusion detection using pseudonyms and provided the motivation for
the work of Biskup et al.\ in \cite{Biskup00a,Biskup00b}. While the
work in \cite{Lundin99,Biskup00a,Biskup00b} does deal with log data
and anonymization, their goals are significantly different than
ours. All three works deal specifically with pseudonymization in
Intrusion Detection Systems (IDSs). The adversary in their model is
the system administrator, and the one requiring protection is the user
of the system. In our case, we instead assume that the system/network
administrators have access to raw logs, and we are trying to protect
the systems from those who would see the shared logs. To contrast how
this makes a difference, consider that in their scenario the server
addresses and services running are not even sensitive---just
information that could identify clients of the system. Furthermore, we
do not require the ability to reverse pseudonyms. However, since the
system/network administrators in their scenario do not have raw data,
the privacy officer must help the system security officer reverse
pseudonyms if alerts indicate suspicious behavior. In
\cite{Biskup00a,Biskup00b}, they take this further and try to support
automatic re-identification if a certain threshold of events is
met. In that way, their pseudonymizer must be intelligent, like an IDS,
predicting when re-identification may be necessary and thus altering
how it pseudonymizes data. They also differ from us in that they
create transactional pseudonyms, so a pseudonym this week might map to
a different entity the next week. We, however, desire consistency with
respect to time for logs to be useful. Lastly, all of the anonymizing
solutions in these papers filter log entries and remove them if they
are not relevant to the IDS; we endeavor to dispose of no entries
because completeness is very important for logs released to the
general research populace. All-in-all, we are looking at the more
general problem of sharing arbitrary logs, rather than hooking
anonymization into IDS's or other tools for very specific purposes.

In \cite{Flegel02}, Flegel takes his previous work in privacy
preserving intrusion detection \cite{Biskup00a,Biskup00b} and changes
the motivation slightly. Here, he imagines a scenario of web servers
volunteering to protect the privacy of visitors from themselves, and
he believes IP addresses of visitors need pseudonymization. However,
to a web server IP addresses already act as a pseudonym protecting the
client's identity, since ISPs rarely volunteer IP-to-person mappings
to non-government entities. Though the motivation differs slightly,
the system described is the same underlying threshold based
pseudonymization system, and the focus of this paper is really about
the implementation and performance of the system. As such, the results
of \cite{Flegel02} can be applied to \cite{Biskup00a,Biskup00b}.

In \cite{Pang03}, Pang et al.\ developed a new packet anonymizer that
anonymizes packet payloads as well as transactional information,
though their methodology only works with application level protocols
that their anonymizer understands: HTTP, FTP, Finger, Ident and
SMTP. The process can also alter logs significantly, losing
fragmentation information, the size and number of packets and
information about retransmissions; thus skewing time stamps, sequence
numbers and checksums.  While their anonymizer is limited in its
capabilities, it is fail-safe because it only leaves information in
the packets that it can parse and understand. Further, they create a
classification of anonymization techniques and a classification of
attacks against anonymization schemes that we found useful. We use a similar
classification which is based off of their work.

Lincoln et al.\ \cite{Lincoln04} proposed a log repository framework that
enables community alert aggregation and correlation, while maintaining
privacy for alert contributors. However, the anonymization scheme in
the paper is partially based on hashing IP addresses.  Such a scheme
is always vulnerable to dictionary attacks. Moreover, their scheme
mixes the hashes with HMACs and truncates the hashes/MACs to 32 bits,
both actions which result in more hash collisions and inconsistent
mappings.  Finally, their suggested use of re-keying by the repository
destroys the correlation between repositories and therefore limits the
view to a single repository.

Most recently, Pang et. al have taken a new approach to packet trace
anonymization, ignoring packet data \cite{Pang06}. Their new tool
anonymizes many fields in pcap logs, more than any other tool to date.
The algorithm options to anonymize these fields are very customized to
their specific needs which they describe through the paper much like a
case study. However, they leave hooks into the software that would
allow someone to create new algorithms to anonymize any TCP/IP fields
in new ways.  The difference between our work is that theirs is (1)
restricted to pcap headers, (2) is not a modular framework with a
clean API, and (3) not many options are available for any particular
field, though the ability for someone else to code another algorithm
is there. Another way to look at the difference is that they just look
at pcap headers and leave open the ability for others to add
anonymization algorithms to their parser.  We create a suite of
anonymization algorithms with a policy manager, and recognizing that
many logs share common fields, we leave open the ability to add new
parsers.

\section{Conclusions and Future Work}\label{conclusion}

As we can see, the ability to share logs can vastly improve the
detection of zero-day exploits and other network attacks. In addition,
the sharing of network traces will allow multiple organizations to
pursue research on real-life examples, instead of simulations. The
main problem with sharing logs is to make sure that sensitive data is
not compromised by sharing the log. But while we should make sure
sensitive information is not distributed, we must also make sure
enough information is retained for analysis. We propose that
anonymization can be used to sanitize the sensitive information in a
log while keeping enough information for analysis. Our goal is to
explore the trade-offs between the strength of an anonymization
algorithm (the amount of information it hides) and its utility (the
usefulness of the log after anonymization). We have developed a
tool---Framework for Log Anonymization and Information Management
(FLAIM)---which we will use to explore these issues.

Log anonymization tools to date have been created in an ad hoc manner
for specific needs of individual organizations.  \flaim is immanently
more flexible than the log anonymization tools to date. FLAIM is
extensible, supports multi-level anonymization with a rich supply of
anonymization algorithms and supports multiple log formats. We have
discussed its goals and architecture in depth, and shown it to perform
admirably at anonymizing large data sets ({\it Preliminary estimates
are GB per minute. The final version of this paper will have exact
numbers isolating anonymization from file I/O}). With the release of
FLAIM ({\it estimated Jul. 2006}), organizations will be able to
leverage our work to address their unique log sharing needs.

\flaim was developed with the purpose of aiding security engineers in
safely sharing security related data with other professionals. While
this is still the main goal of FLAIM, it is important to note that
\flaim can be used on any type of data, provided a module is written
for it. We have seen that there is a need for anonymization to allow
the further progress of science. Because of FLAIM's separation of the
file I/O from the actual anonymization, \flaim can provide a
general framework for anonymization of any data. Any developer can
make use of the diverse set of anonymization algorithms we provide to
anonymize their data. By releasing \flaim as an open source tool, and
by creating a simple module API, we hope that many professionals will
build modules for FLAIM.

While FLAIM is a capable tool, there are always enhancements that can
be made.  New parsing modules may be written for parsing additional
log types. These parsing modules can also be enhanced to support
streaming or real-time anonymization.  We also foresee interest in a
daemon mode for FLAIM that would make it easier to integrate with web
services and other online tools. Most interesting to us, is the
creation of a tool to help generate anonymization policies. It is
always challenging to determine how best to anonymize the data for a
particular application.  We envision a tool that asks users a series
of questions to determine how to anonymize a log, and then it would
output a valid XML policy for FLAIM. As we receive feedback from
users, we expect to have more solid grasp of which features are most
useful.

\section{Acknowledgements}
This material is based in-part upon work supported by the National
Science Foundation under award No. CNS-0524643.

\end{document}